%% file: main.tex
\documentclass[runningheads]{llncs}
\usepackage{graphicx}
\usepackage[utf8]{inputenc}
\usepackage[T1]{fontenc}
\usepackage[separate-uncertainty = true, multi-part-units = repeat]{siunitx}
\usepackage{amsmath,bm}
\usepackage{float}
\begin{document}
\title{Quantitative Hemodynamics in Aortic Dissection: Comparing \textit{in vitro} MRI with FSI Simulation in a Compliant Model.}
\titlerunning{Hemodynamics in TBAD: FSI vs. 4D-Flow}
\author{Judith Zimmermann \inst{1,2}\orcidID{0000-0002-9579-9788} 
		\and
		Kathrin B\"aumler \inst{1}\orcidID{0000-0002-2126-5919}
		\and
		Michael Loecher  \inst{1,3}\and
		Tyler E. Cork  \inst{1}\orcidID{0000-0001-5759-5322}\and
		Fikunwa O. Kolawole \inst{5}\and
		Kyle Gifford \inst{1}\and
		Alison L. Marsden  \inst{4}\and
		Dominik Fleischmann  \inst{1}\and
		Daniel B. Ennis\inst{1,3}}
\authorrunning{J. Zimmermann et al.}

\institute{Department of Radiology, Stanford University, USA \and
	Department of Computer Science, Technical University of Munich, Germany \and
	Division of Radiology, VA Palo Alto Health Care System, USA \and
	Department of Pediatrics, Stanford University, USA \and
	Department of Mechanical Engineering, Stanford University, USA}
\maketitle              
\begin{abstract}
The analysis of quantitative hemodynamics and luminal pressure may add valuable information to aid treatment strategies and prognosis for aortic dissections. This work directly compared \textit{in vitro} 4D-flow magnetic resonance imaging (MRI), catheter-based pressure measurements, and computational fluid dynamics that integrated fluid-structure interaction (CFD FSI). Experimental data was acquired with a compliant 3D-printed model of a type-B aortic dissection (TBAD) that was embedded into a flow circuit with tunable boundary conditions. \textit{In vitro} flow and relative pressure information were used to tune the CFD FSI Windkessel boundary conditions. Results showed overall agreement of complex flow patterns, true to false lumen flow splits, and pressure distribution. This work demonstrates feasibility of a tunable experimental setup that integrates a patient-specific compliant model and provides a test bed for exploring critical imaging and modeling parameters that ultimately may improve the prognosis for patients with aortic dissections.
\end{abstract}
\input{introduction}
\input{methods}
\input{results}
\input{discussion}
\subsection*{Acknowledgements}
We thank the Stanford Research Computing Center for computational resources (Sherlock HPC cluster), Dr. Anja Hennemuth for making available software tools, and Nicole Schiavone for technical advice. Funding was received from DAAD scholarship program (to J.Z.) and NIH R01 HL13182 (to D.B.E).
%
% ---- Bibliography ----
\bibliographystyle{splncs04}
\bibliography{library}
\end{document}

%% file: introduction.tex
\section{Introduction}
An aortic dissection is a life-threatening vascular disorder in which a focal tear develops within the inner aortic wall layer. This leads to subsequent formation of a secondary channel (`false lumen', FL) that is separated from the primary channel (`true lumen', TL) by a dissection flap.~\cite{Nienaber2016}
Patients with type-B aortic dissection (TBAD, i.e. without involvement of the ascending aorta) often receive pharmacologic treatment and frequent monitoring is used in an attempt to predict late adverse events. Prognosis of late adverse events is largely informed by morphologic imaging features, but conflicting results have been reported among several predictors~\cite{Spinelli2018}. 

To improve prognosis several hemodynamic quantities are of potential interest and may confer added sensitivity of individual risk. Recent studies have suggested high FL outflow~\cite{Sailer2017} as strong predictor for late adverse events, and FL ejection fraction~\cite{Burris2020} as indicator for false lumen growth rate.

To retrieve these hemodynamic markers, computational fluid dynamics (CFD) frameworks provide simulated patient-specific flow fields at high spatio-temporal resolution~\cite{Marsden2013}; 
and those that integrate fluid-structure interaction (FSI) at deformable walls are expected to amplify the realism of patient-specific modeling even further. 
If simulations were able to reliably replicate hemodynamics, it would further enable non-invasive prediction of risk related to pathological changes (e.g. tear size).

While CFD FSI approaches show great potential, a direct validation with measured data in highly controlled, but realistic environments is missing. Previous comparisons between simulations and \textit{in vivo} 4D-flow MRI~\cite{Baumler2020,Dillon-Murphy2016,Pirola2019} are challenged by: the assumption of a rigid aortic wall and dissection flap; a lack of information on accurate patient-specific hemodynamic conditions; and/or an unknown patient-specific aortic wall and dissection flap compliance.

Herein, we compare qualitative and quantitative TBAD hemodynamics based on: (1) simulations that use a recently proposed FSI framework~\cite{Baumler2020}, and (2) \textit{in vitro} MRI including catheter-based pressure mapping. We utilized a patient-specific, compliant TBAD model embedded into a highly-controlled flow circuit. Uniaxial tensile testing of the compliant material, image-based flow splits and catheter-based pressure recordings informed simulation tuning.

%% file: methods.tex
\section{Methods}

\subsection{Patient-specific aortic dissection model} \label{Sec:2.1}

A 3D computed tomography angiogram (CTA) of a patient (31 y/o, female) with TBAD was selected from our institution's database. A proximal intimal `entry' tear was present distal to the left subclavian artery and an `exit' tear was located proximal to the celiac trunk. Each tear measured \SI{2.3}{\square\centi\meter} in area size. The CTA study was approved by the institutional review board and written consent was obtained prior to imaging. 

\begin{figure}[h]
	\centering
	\includegraphics[scale=1]{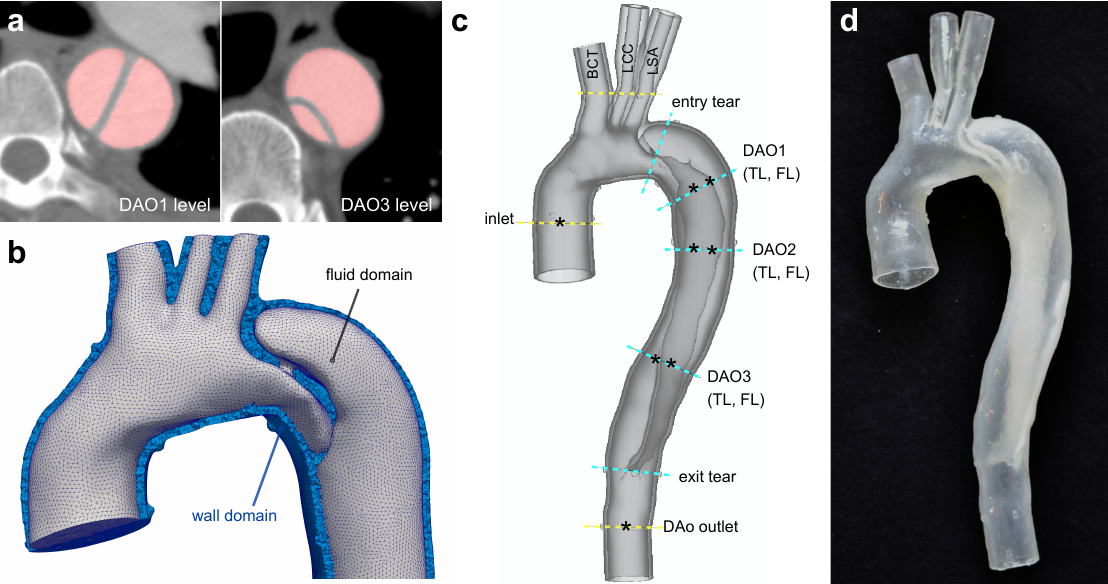}
	\caption{
		(a) CTA images with lumen segmentation. 
		(b) Tetrahedral meshes of fluid (gray) and wall domains with dissection flap (blue). 
		(c) Cross-sectional landmarks and pressure mapping points (*). `Entry' and `exit' tear cover sections with combined TL and FL flow. Landmarks DAO1, DAO2, and DAO3 consist of a TL and FL cross-section. 
		(d) Photograph of finished 3D-printed model.}
	\label{FIG_TBADmodel}
\end{figure}

The lumen of the thoracic aorta  was segmented using the active contour algorithm with supplementary manual refinements (itk-SNAP v3.4, Fig. \ref{FIG_TBADmodel}a). Two tetrahedral meshes were generated (Fig. \ref{FIG_TBADmodel}b): the `fluid domain' representing the full aortic lumen; and the `wall domain' (as extruded fluid domain) representing the outer aortic wall and dissection flap that separates TL and FL with uniform thickness ($h=\SI{2}{\milli\meter}$). The wall domain mesh was further refined with (i) cylindrical caps that facilitated tubing connections, and (ii) visual landmarks to define image analysis planes. Meshing and refinements were done using SimVascular (release 2020-04)~\cite{Updegrove2017} 
and Meshmixer (v3.5, Autodesk). Further details on model generation are given in~\cite{Baumler2020}.

The wall model was 3D-printed using a novel photopolymer technique (PolyJet J735, Stratasys Inc.), as shown in Fig. \ref{FIG_TBADmodel}d. The print material underwent uniaxial tensile testing as described in~\cite{Zimmermann2021c} and proved to be comparable to \textit{in vivo} aortic wall compliance (tangent Young's modulus $E_{y,t} = \SI{1.3}{\mega\pascal}$).

\subsection{MRI experiments}
\begin{figure}[ht]
	\centering
	\includegraphics[scale=0.8]{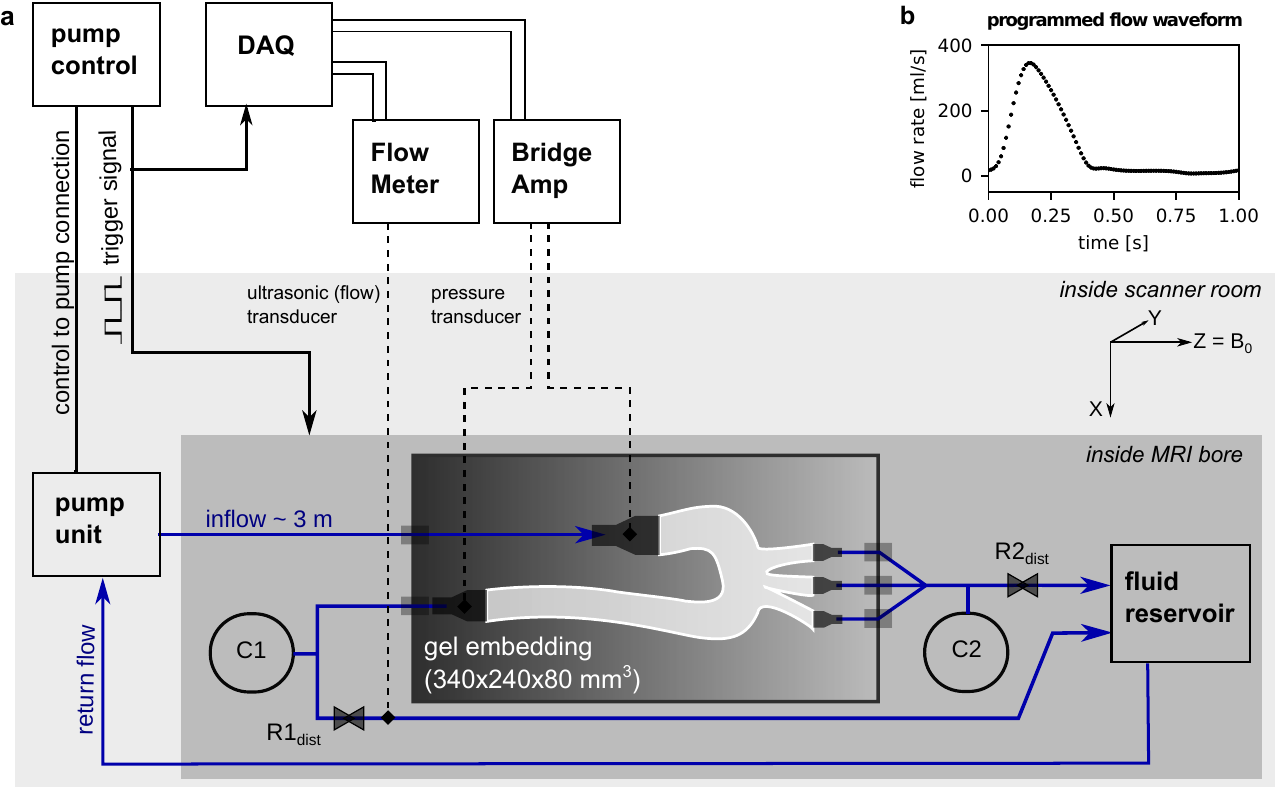}
	\caption{
		(a) Schematic of the flow circuit setup. Pressure transducers were inserted through ports at the model inlet and DAo outlet. (b) Flow rate waveform that was programmed into the pump. $C1, C2$: 'capacitance' air-compression chambers, $R1, R2$: 'resistance' flow clamps.}
	\label{FIG_setup}
\end{figure}

Imaging was performed on a \SI{3}{\tesla} MRI machine (Skyra, Siemens). An MRI-compatible flow circuit that includes a programmable pump (CardioFlow 5000 MR, Shelley) was engineered to provide controllable flow and pressure conditions similar with target values within the physiological range (Fig. \ref{FIG_setup}a). Details were recently published in~\cite{Zimmermann2021c}.
Glycerol-water (ratio = \SI{40}{\percent}/\SI{60}{\percent}) with contrast (ferumoxytol) was used as a blood-mimicking fluid; and a typical aortic flow waveform (Fig. \ref{FIG_setup}b) was applied (heart rate = \SI[per-mode=symbol]{60}{\per\minute}, stroke volume = \SI[per-mode=symbol]{74.1}{\milli\liter\per\second}, total flow = \SI[per-mode=symbol]{4.45}{\liter\per\minute}).

The circuit was tuned on the scanner table prior to image acquisition, targeting a flow split of \SI{70}{\percent}/\SI{30}{\percent} (DAo outlet vs. arch branches), and luminal systolic pressure (at model inlet) of \SI{120}{mmHg}.
The pulse pressure was contolled via capacitance elements---designed as sealed air compression chambers---at the DAo outlet ($C1$) and at the merged arch branches ($C2$). 
A pressure transducer (SPR-350S, Millar) was inserted through ports at the model inlet and DAo outlet, and luminal pressures were recorded at eight points (Fig. \ref{FIG_TBADmodel}c). 
Ultrasonic flow and pressure signals were fed into PowerLab (ADInstruments) for analysis.

\subsubsection*{2D-cine and 2D-PC MRI}
Two-dimensional (2D) acquisitions at landmarks (Fig. \ref{FIG_TBADmodel}c) included: 
(1) 2D cine gradient echo (2D-cine) with pixel size = \num{0.9x0.9} \si{\square\milli\meter}, 
slice thickness = \SI{6}{\milli\meter}, 
$T_E/T_R$ = \num{3/4.75} \si{\milli\second}, 
flip angle = \ang{7}, 
avg. = 2, 
retro. gating (40 frames); and 
(2) 2D phase-contrast (2D-PC) with pixel size = \num{1.1x1.1} \si{\square\milli\meter}, 
slice thickness = \SI{6}{\milli\meter}, 
$T_E/T_R$ = \num{3/5.25} \si{\milli\second}, 
flip angle = \ang{25}, 
avg. = 2,  
$V_{enc}$ = 90--120 \si{\centi\meter\per\second}, 
retro. gating (40 frames). 

\subsubsection*{4D-flow MRI}
A four-point encoded Cartesian 4D-flow sequence was acquired as follows: FoV = \num{340x236x84} \si{\cubic\milli\meter}, matrix = \num{220x156x56}, voxel size = \num{1.5x1.5x1.5} \si{\cubic\milli\meter}, $T_E/T_R$ = \num{2.7/5.6} \si{\milli\second}, flip angle = \ang{15}, parallel imaging (GRAPPA, R=2), $V_{enc}$ = \SI{120}{\centi\meter\per\second}, lines/seg. = 2, retro. gating (20 frames).

\subsubsection*{Image analysis}
Lumen contours were automatically tracked through time based on 2D-cine data via image-based deformable registration,
which provided values of cross-sectional area and served as the boundary for net flow calculation.
2D-PC images were corrected for phase offsets (via planar $2^{nd}$ order fitting) and then processed to retrieve the inlet flow and net flow splits across outlets.

4D-flow MRI data was corrected for (i) Maxwell terms, (ii) gradient non-linearity distortion~\cite{Markl2003},
and (iii) phase offsets (via $3^{rd}$ order fitting). Five landmarks along the dissected region were used for analysis (Fig. \ref{FIG_TBADmodel}c). 4D-flow MRI offset correction, flow calculations, and streamline visualization were done using MEVISFlow (v11.2, Fraunhofer Institute) and ParaView (v5.7); quantitative results were exported as numeric files for comparison with simulation results.

\subsection{CFD FSI simulations}
\subsubsection{Governing Equations} 
The governing equations for fluid flow and structural mechanics were solved in the fluid and wall domain, respectively.
In the fluid domain, the working fluid was considered incompressible and Newtonian ($\varrho_f = \SI{1100}{\kilogram\per\cubic\meter}$, $\mu_f = \SI{0.00392}{\pascal\second}$). Momentum and mass balance were described by the Navier-Stokes Equations in arbitrary Lagrangian Eulerian formulation to account for motion.
The structural material was modeled with a Neo-Hookean model for homogeneous, isotropic hyperelastic materials ($E_{y} = \SI{1.3}{\mega\pascal}$, $\varrho_s = \SI{1450}{\kilogram/\meter^3}$). Both domains were coupled at the interface via kinetic and dynamic interface conditions. A detailed mathematical description can be found in~\cite{Baumler2020}.

\subsubsection{CFD FSI boundary conditions}
%Three-element Windkessel boundary conditions were applied at fluid outlets and coupled to the 3D domain with the coupled multidomain method~\cite{Esmaily-Moghadam2013a}. A distal and proximal resistance and capacitance was determined for each outlet in an iterative approach via manually tuning the Windkessel parameters. The catheter-based pressure values used as simulation tuning targets were: $\SI{119}{\mmHg}$, $\SI{42}{\mmHg}$, and $\SI{77}{\mmHg}$ for the systolic ($P_s$), diastolic ($P_d$), and pulse pressure ($\triangle P$), respectively. Tuning was stopped once a \SI{10}{\percent} error margin was reached. The 2D-PC derived flow waveform was prescribed at the model inlet, assuming a parabolic velocity profile. 2D-PC flow splits (w.r.t. inlet flow) were prescribed with \SI{79.4}{\percent}, \SI{12.3}{\percent}, \SI{3.1}{\percent}, and \SI{5.3}{\percent} for the DAo outlet, BCT, LCC, and LSA, respectively.

The 2D-PC derived flow waveform was prescribed at the model inlet as a Dirichlet boundary condition, assuming a parabolic velocity profile.  Three-element Windkessel boundary conditions were applied at fluid outlets and coupled to the 3D domain with the coupled multidomain method~\cite{Esmaily-Moghadam2013a}.  The catheter-based pressure values at the inlet of the model used as simulation tuning targets were: $\SI{119}{\mmHg}$, $\SI{42}{\mmHg}$, and $\SI{77}{\mmHg}$ for the systolic ($P_s$), diastolic ($P_d$) and pulse pressure ($\triangle P$), respectively. Additionally, the 2D-PC derived flow splits informed the Windkessel parameter tuning, and were measured as \SI{78.4}{\percent}, \SI{12.3}{\percent}, \SI{3.0}{\percent}, and \SI{5.2}{\percent} for the DAo outlet, BCT, LCC, and LSA, respectively. 
The tuning of the Windkessel parameters (a distal and proximal resistance and capacitance at each of the model outlets) was then carried out in an iterative and manual process, by which a total resistance $R_T$ and total capacitance $C_T$ are distributed across all model outlets according to the measured flowsplits and a pre-prescribed ratio of distal to proximal resistance ($k_d = 0.9$). Details of the tuning process can be found in~\cite{Baumler2020}.

Wall domain outlets were fixed in space via a homogeneous Dirichlet condition for the displacement and a homogeneous Neumann boundary condition was prescribed at the outer wall of the vessel domain. This is in contrast to patient-specific simulations where a non-homogeneous Robin type boundary condition can be prescribed to account for external tissue support of the vessel. Likewise, the outer wall of the vessel domain was assumed to not be under prestress, contrary to a typical \textit{in vivo} environment.

\subsubsection{Numerical formulation}
The numerical simulations were performed with the SVFSI finite element solver, as implemented in SimVascular~\cite{Updegrove2017}. 
SVFSI features linear elements for velocity and pressure and is based on the ''Residual Based Variational Multiscale`` method. The fluid and wall domain were solved in a monolithic approach and backflow stabilization was applied at the fluid outlets. To avoid mesh degeneration, a nodal mesh smoothing was performed after each time step. 
Details of the numerical formulation are given in~\cite{Baumler2020}. For details about the numerical discretization we refer to~\cite{Bazilevs2007,Bazilevs2008,Tezduyar1992,Esmaily-Moghadam2015}.

\subsubsection{Discretization and simulation setup}
Tetrahedral meshes of fluid and wall domain were sampled with a
spatial resolution of $\triangle h=\SI{1.3}{\milli\meter}$ ($\num{1.6e6}$ tetrahedral elements) which was found to be a sufficiently fine resolution~\cite{Baumler2020}. The temporal resolution was set to \num{4e3} timesteps per cardiac cycle ($\triangle t = \SI{0.25}{\milli\second}$). The simulation achieved cycle-to-cycle periodicity within \num{5} iterative runs. Compute time was $\SI{12}{\hour}$ per cycle on a high performance computing cluster.

\subsubsection{CFD FSI analysis}
Time-resolved parameters were extracted from the last simulation cycle:
(i) flow rate, (ii) area change, and (iii) pressure. We extracted data from every 50th simulated time step, which totaled 80 incremental results with an effective temporal resolution $\triangle t = \SI{12.5}{\milli\second}$. Quantitative metrics were analyzed at cross-sectional landmarks (Fig. \ref{FIG_TBADmodel}c) using ParaView (v5.7) and exported as numeric files for direct 4D-Flow MRI comparison.

%% file: results.tex
\section{Results}
\subsubsection{Boundary conditions}
\begin{figure}[t]
	\centering
	\includegraphics[scale=1]{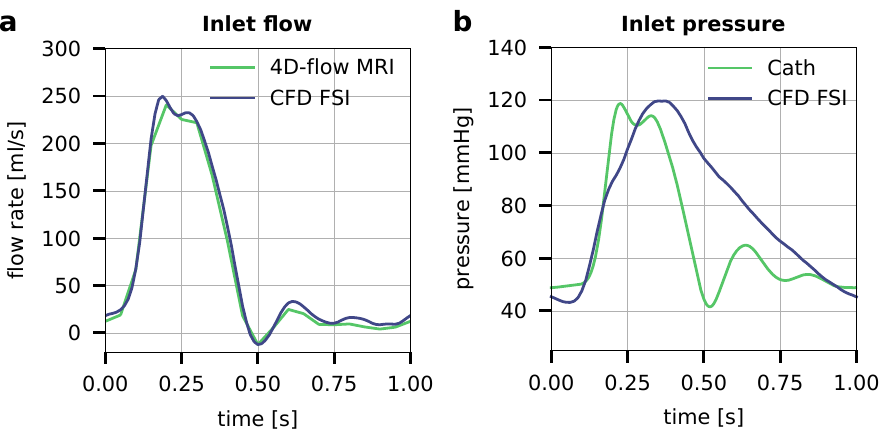}
	\caption{CFD FSI (blue) tuning conditions, showing (a) flow rate and (b) pressure waveform at TBAD model inlet, in comparison to 4D-flow and catheter measurements (green). 
	While the inlet flow rate waveforms match well, CFD FSI shows a slower diastolic pressure decay without oscillations.
	}
	\label{FIG_tuningCond}
\end{figure}

Inlet flow (Fig. \ref{FIG_tuningCond}a) for CFD FSI was directly prescribed based on 2D-PC results and agreed well with 4D-flow MRI. CFD-FSI flow splits across model outlets 
(\SI{78.7}{\percent},
\SI{12.7}{\percent},
\SI{3.2}{\percent}, and
\SI{5.5}{\percent}
for DAo outlet, BCT, LCC, and LSA, respectively) 
aligned well with 2D-PC splits 
(\SI{78.4}{\percent},
\SI{12.1}{\percent},
\SI{3.0}{\percent}, and
\SI{5.2}{\percent}). 
After tuning, CFD FSI pressure (Fig. \ref{FIG_tuningCond}b) matched catheter measurements within the pre-defined \SI{10}{\percent} error margin 
(\SI{119.6}{\mmHg},
\SI{43.2}{\mmHg}, and
\SI{76.4}{\mmHg} for simulated $P_s$, $P_d$ and $\triangle P$, corresponding to a relative error of $\SI{\leq 4}{\percent}$). 
While catheter-based measurements showed oscillations and a fast pressure drop at end-systole ($t=\SI{0.4}{\second}$), CFD FSI pressure decayed slower and without oscillation. As a results, mean pressure differed by \SI{15.8}{\percent} 
(\SI{78}{\mmHg} for CFD FSI compared to \SI{68}{\mmHg} for catheter-based measurements).
\subsubsection{Flow patterns and velocities}
\begin{figure}[t]
	\centering
	\includegraphics[scale=0.85]{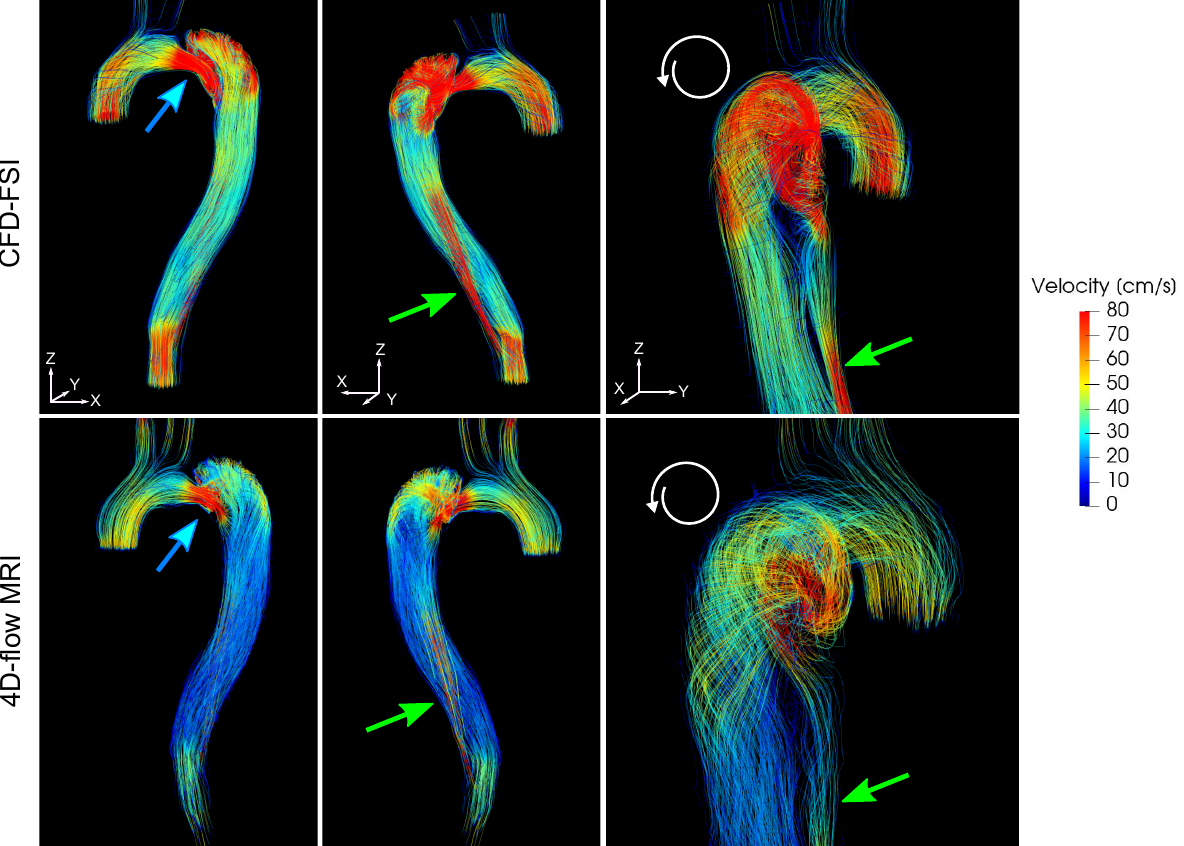}
	\caption{Streamline visualization at systole ($t=\SI{0.2}{\second}$) for CFD FSI (top) and 4D-flow MRI (bottom) data. 
		CFD FSI shows higher velocities, but intra-modality flow patterns and velocity distribution is consistent. Increased velocities through entry tear (blue arrows) and true lumen (green arrows). A helical flow pattern is visible in the false lumen aneurysm (white arrows).
	}
	\label{FIG_streamlines}
\end{figure}

Qualitative flow visualizations (Fig. \ref{FIG_streamlines}) showed well-matched flow patterns between CFD FSI and 4D-flow MRI. Particularly, streamlines depicted helical flow in FL aneurysm during systole and distal FL during diastole, as well as increased velocities through the proximal FL entry tear and along the distal TL. Overall, velocities were higher in CFD FSI, but the intra-model spatial distribution of velocities matched well.
\subsubsection{Pressure, area, and flow}
\begin{figure}[t]
	\centering
	\includegraphics[width=\textwidth]{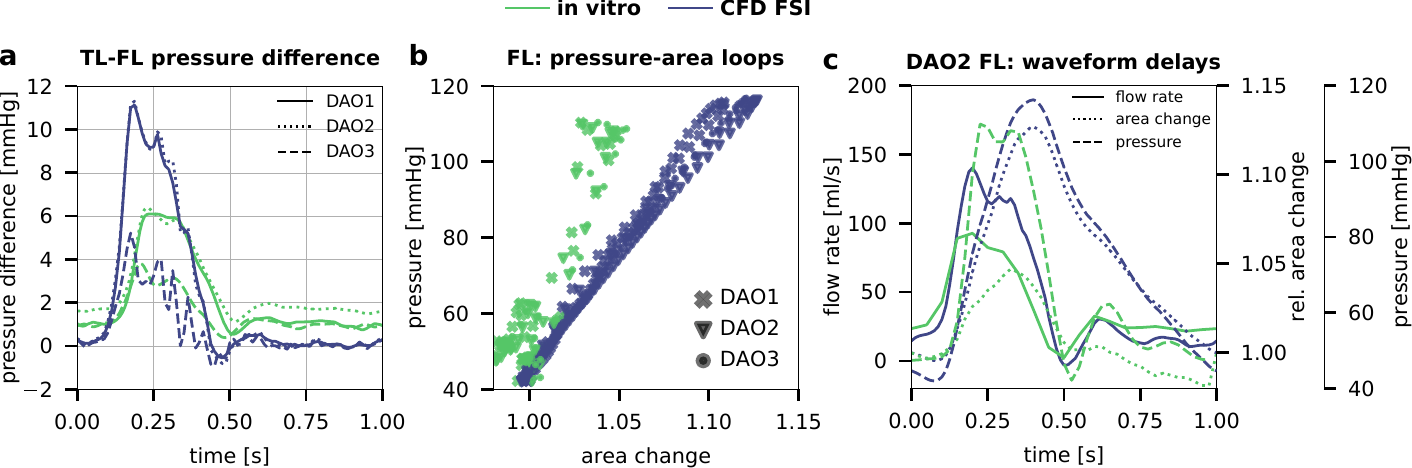}
	\caption{(a) The TL-FL pressure difference was higher in proximal and lower in distal region. 
	(b) FL pressure-area loops. 
	(c) Flow rate peaks preceding both pressure and area peaks, with greater delay times for CFD FSI.}
	\label{FIG_results_pressure}
\end{figure}

\begin{figure}[]
	\centering
	\includegraphics[scale=1]{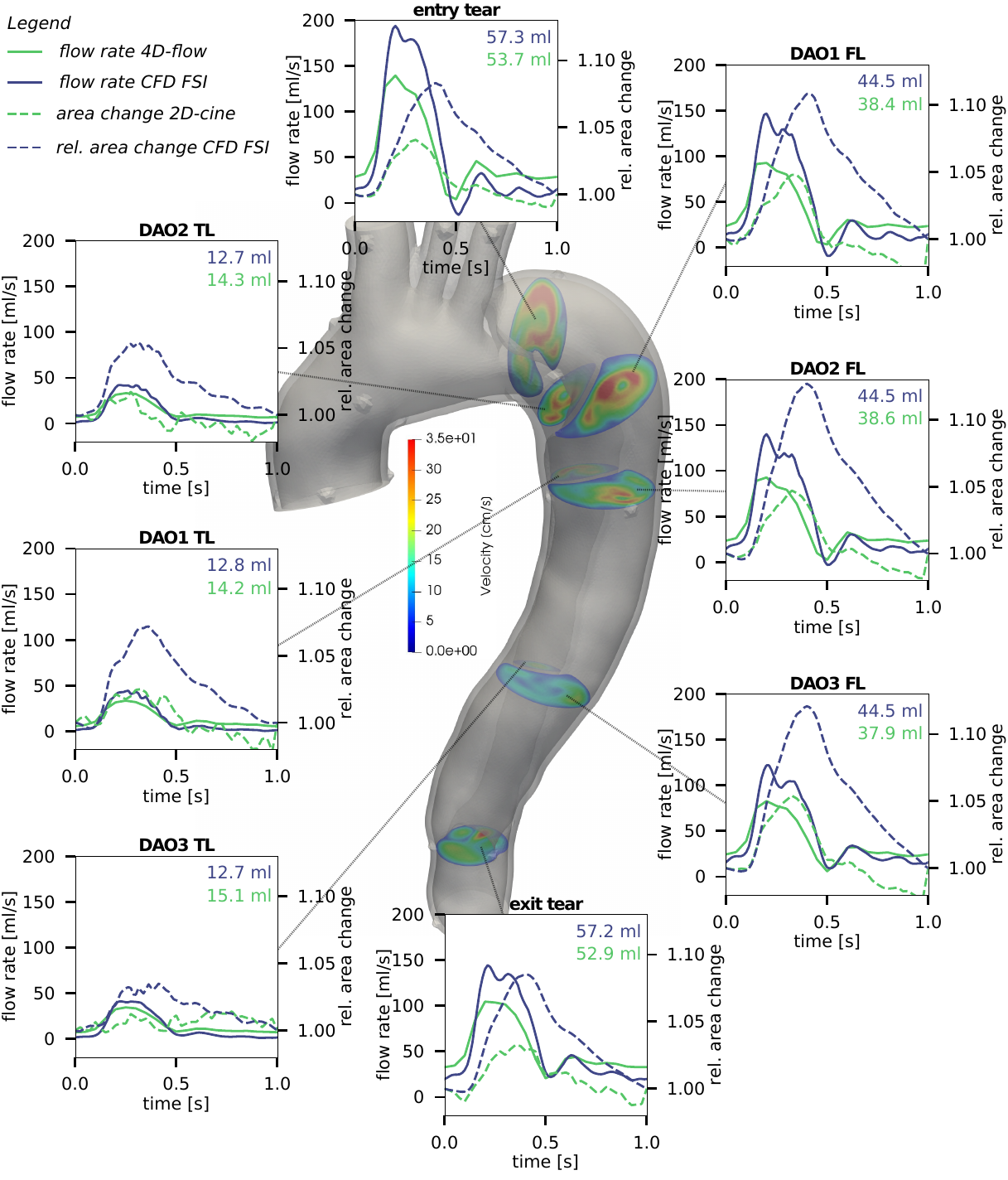}
	\caption{Flow rate and area change (w.r.t area of first frame) at eight landmarks. Net flow values for CFD FSI (blue) and 4D-flow MRI (green) are given. 
	CFD-FSI showed increased FL flow (\SI{78}{\percent}) compared to 4D-flow (\SI{73}{\percent}); and increased maximum area expansion (\SI{11}{\percent} for CFD FSI vs. \SI{5}{\percent} for 4D-flow).
	}
	\label{FIG_flowrate_area}
\end{figure}

Systolic TL pressure exceeded FL pressure (Fig. \ref{FIG_results_pressure}a) for both simulation and catheter measurements. At peak systole, the TL-FL presure difference was greater for CFD FSI data at landmarks DAO1 and DAO2, but matched well at DAO3. During diastole, the TL-FL difference was close to zero for CFD FSI, but was 1 to 2 \si{\mmHg} for the catheter measurements. 
Cross-sectional area (Fig. \ref{FIG_flowrate_area}, dashed lines) expanded most in FL cross-sections with up to \SI{11}{\percent} based on CFD FSI and up to \SI{5}{\percent} based on 2D-cine MRI. 

Net flow volumes (Fig. \ref{FIG_flowrate_area}) revealed a FL to TL flow split of \SI{78}{\percent}/\SI{22}{\percent} for CFD FSI and \SI{73}{\percent}/\SI{27}{\percent} for 4D-flow MRI measurements. Flow waveform shapes (Fig. \ref{FIG_flowrate_area}, solid lines) aligned well, particularly regarding the peak flow timepoint, systolic upslope ($t = \SI{0.1}{\second}$), and oscillatory lobes in diastole. CFD FSI flow rates were higher in systole and lower in diastole when compared to 4D-flow values. 

Pressure-area loops showed a steeper slope for \textit{in vitro} data (Fig. \ref{FIG_results_pressure}b). FL peak flow preceded peaks of pressure and area change. This temporal delay was longer for CFD FSI, which was consistent for all DAO landmarks (Fig. \ref{FIG_results_pressure}c). 

\begin{figure}[t]
	\centering
	\includegraphics[scale=1]{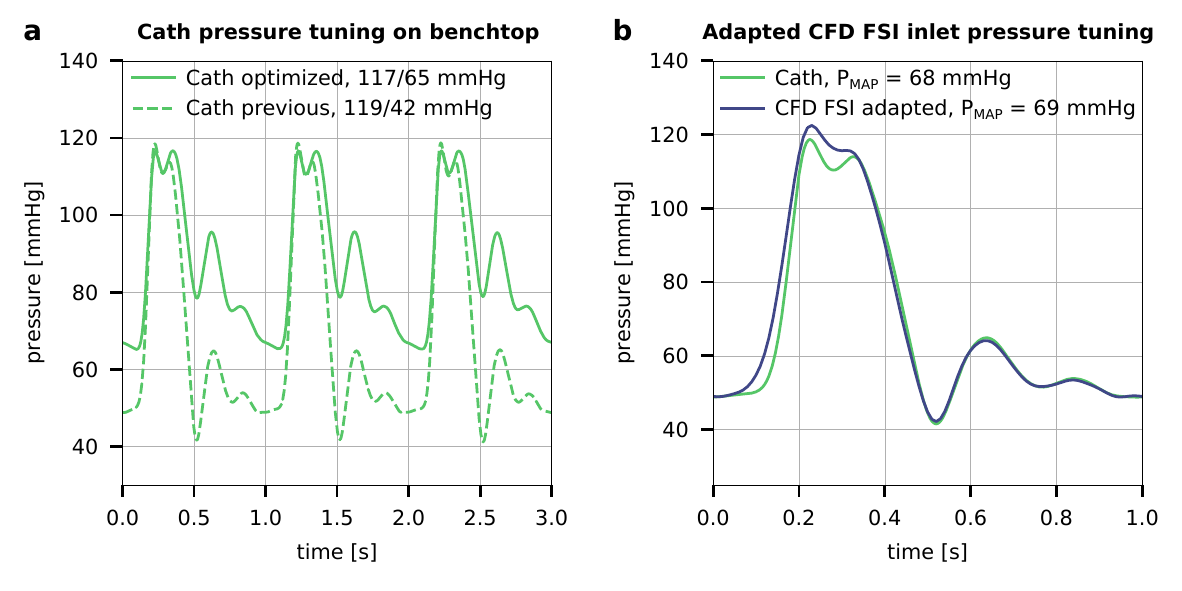}
	\caption{
		(a) Exploratory benchtop data showing improvements for the setup's pressure condition. By drastically reducing proximal resistance, pulse pressure is reduced from \SIrange[range-units=single]{78}{52}{\mmHg}, and --- if wave reflections were to be neglected --- diastolic pressure declines steadily towards its minimum at end-diastole. (b) CFD FSI (blue) vs catheter-based (green) pressure at the inlet face in an adapted simulation with rigid walls and inhomogeneous pressure boundary condition at the BCT outlet. By tuning the simulation using the full pressure waveform, inlet pressure conditions closely resemble measured values, including wave reflections.}
	\label{FIG_pressure_pre_post_k_d}
\end{figure}

%% file: discussion.tex
\section{Discussion}
This study leveraged compliant 3D-printing as well as a highly-controlled MRI-compatible flow circuit setup to directly compare CFD FSI and MRI results with regards to flow and pressure dynamics in a patient-specific TBAD model. The aorta's secondary lumen and proximal FL aneurysm presented complex flow patterns with a large velocity range. These characteristics were well captured by both modalities and streamline visualizations were in very good agreement. 

Our approach links measured luminal pressure with CFD FSI boundary conditions, which presents a major advantage over previous comparisons with \textit{in vivo} data that usually lack invasive pressure measurements. During simulation tuning, pressure targets ($P_s$, $P_d$) were met, but pressure waveform shapes differed---i.e. faster and oscillatory pressure decay in catheter measurements versus slower and steady decline in CFD FSI. We note that a slower and steady pressure decline in diastole is desirable and would resemble \textit{in vivo} pressure shapes of the arterial system~\cite{Mills1970}.

To further investigate this mis-match, additional exploratory pressure data were recorded on the benchtop. Moreover, additional CFD FSI simulations with varying configurations of boundary conditions were computed. 
We identified three aspects to better match the measured and simulated pressure conditions. 
First, increasing the ratio between the distal and proximal resistance---described by parameter $k_d$ in the three-element Windkessel model---is the key factor to improve the pressure shape towards a slower diastolic decline with its minimum at end-diastole (Fig. \ref{FIG_pressure_pre_post_k_d}a). In practice, we increased $k_d$ by decreasing proximal resistance, and thus also decreased total system resistance which led to reduced $\triangle P$.
Second, exploratory benchtop experiments also suggested that the characteristic pressure oscillations originate from wave reflections at multiple branching points. Other previous works that deployed flow circuits in model-based studies reported similar pressure waveform shapes~\cite{Birjiniuk2017,Gallarello2019,Urbina2016b,Schiavone2021}. Further engineering efforts should be made to minimize wave reflections at non-smooth boundaries as much as possible.
Third, additional exploratory simulation runs suggested that the measured pressure conditions can be closely approximated by directly prescribing pressure data as inhomogeneous Neumann boundary condition at one of the outlets. To do so, we prescribed the catheter-based pressure data that was measured at the model inlet as pressure boundary condition at the brachiocephalic trunk outlet. Resulting well-matched pressure waveforms are shown in Fig. \ref{FIG_pressure_pre_post_k_d}b. We seek to adapt both our experimental and simulation setup in future studies regarding these three aspects.

Overall, our presented results showed similar tendencies of flow and pressure parameters in the dissected region between MRI, catheter measurements, and FSI simulation. TL-FL pressure differences were comparable such that they were almost consistently positive, and that the most distal landmark (DAO3) showed a smaller difference compared to the two proximal points (DAO1, DAO2). Interestingly, CFD FSI TL-FL pressure difference briefly dropped to negative ($t=\SI{0.4}{\second}$) and then to zero ($t>\SI{0.65}{\second}$), while catheter measurements showed preserved positive TL-FL differences at all locations and times. Moreover, with only \SI{5}{\percent} difference between modalities, results suggest a well-matched TL-FL flow split.

Multiple results indicate that the performed tensile testing underestimated $E_{y, t}$: a steeper slope of the pressure-area loop for \textit{in vitro} data, shorter flow-pressure-area waveform delays, and consistently lower outer wall expansion. To address this mis-match, future CFD FSI experiments should iteratively increase the value for $E_{y, t}$ until \textit{in vitro} wall deformations are sufficiently replicated. 

Moving toward clinical deployment of simulation-based treatment decision support, future work should also investigate uncertainties of pressure and flow conditions and their impact on hemodynamic quantities. In particular, if pressure data are unavailable, it should be investigated how approximations of pressure boundary conditions (e.g. two-point systolic to diastolic cuff pressure) propagate errors into hemodynamic quantities. The presented highly-controlled \textit{in vitro} setup is well suited to investigate these effects.

%We note that this study is limited by only evaluating one TBAD model, which restricts the generalization of the results. We seek to explore multiple patient models as well as alterations in entry and exit tear size in our future work. 

In conclusion, this work presents valuable information on hemodynamic similarities and differences as retrieved from CFD FSI, \textit{in vitro} MRI, and catheter-based pressure measurements in a patient-specific aortic dissection model.  